\documentclass[%
 reprint,
 amsmath,amssymb,
 aps,
prb,
floatfix,
]{revtex4-2}

\usepackage[version=4]{mhchem}
\usepackage{mathtools}
\usepackage{csquotes}
\usepackage{graphicx}
\usepackage{dcolumn}
\usepackage{bm}
\usepackage{color}
\usepackage[table,xcdraw,dvipsnames]{xcolor}
\usepackage{multirow}
\usepackage{array}
\usepackage{booktabs}
\usepackage{siunitx}[=v2]
\usepackage[breaklinks]{hyperref}

\hypersetup{colorlinks=true, linkcolor=blue, citecolor=blue, filecolor=blue, urlcolor=blue}

\DeclareSIUnit{\angstrom}{\textup{\AA}}

\newcommand{\figph}[2][\linewidth]{%
  \IfFileExists{#2}%
    {\includegraphics[width=#1]{#2}}%
    {{\setlength{\fboxsep}{0pt}\fbox{\parbox[c][0.28\textheight][c]{\dimexpr#1-2\fboxrule\relax}{\centering\ttfamily\footnotesize[figure placeholder]}}}}%
}

\DeclareSIUnit\atom{atom}

\DeclarePairedDelimiter{\ket}{\lvert}{\rangle}
\DeclarePairedDelimiter{\aqty}{\langle}{\rangle}
\DeclarePairedDelimiter{\pqty}{(}{)}
\DeclarePairedDelimiter{\bqty}{[}{]}

\DeclarePairedDelimiterX{\braket}[2]{\langle}{\rangle}{#1\,\delimsize\vert\,\mathopen{}#2}
\DeclarePairedDelimiterX{\mel}[3]{\langle}{\rangle}{#1\,\delimsize\vert\,\mathopen{}#2\,\delimsize\vert\,\mathopen{}#3}

\begin{document}

\preprint{APS/123-QED}

\title{Symmetry-based modal analysis of heat transport in molecular dynamics of quasi-1D systems}

\author{Yu-Jie Cen}
\affiliation{Institute of Materials Chemistry, TU Wien, A-1060 Vienna, Austria}

\author{Sandro Wieser}
\affiliation{Institute of Materials Chemistry, TU Wien, A-1060 Vienna, Austria}

\author{Georg K. H. Madsen}
\email{georg.madsen@tuwien.ac.at}
\affiliation{Institute of Materials Chemistry, TU Wien, A-1060 Vienna, Austria}

\author{Jesús Carrete}
\email{jcarrete@unizar.es}
\affiliation{Instituto de Nanociencia y Materiales de Aragón, CSIC-Universidad de Zaragoza, 50009 Zaragoza, Spain}

\begin{abstract}
Detailed analysis of thermal conductivity results obtained from molecular dynamics (MD) trajectories conventionally relies on knowledge of the harmonic vibrational modes of the system. This is the case in methods like Green--Kubo modal analysis (GKMA) and homogeneous nonequilibrium modal analysis (HNEMA).  However, arbitrary mixing within degenerate phonon subspaces makes individual modal contributions basis dependent and can obscure their symmetry character. We propose an alternative for quasi-one-dimensional (quasi-1D) systems: we construct the modal basis for the decomposition of the thermal conductivity from line-group projection operators, so that every projected component carries well-defined symmetry labels (including rotational information and parities) and the decomposition is unique at the level of irreducible representations (irreps). Applying the idea to a $(10,0)$--$(20,0)$ \ce{WS2}--\ce{MoS2} double-walled nanotube (DWNT) described by a neuroevolution potential (NEP), we find that at \SI{300}{\kelvin} both HNEMA and GKMA yield statistically consistent total conductivities and allow the identification of several prominent symmetry-adapted conduction channels. The GKMA pair matrix shows that within-block and same-channel cross-$k$ terms account for $75.6\%$ of the total conductivity, while cross-channel correlations contribute $24.4\%$.
\end{abstract}

\maketitle


\section{Introduction}

Symmetry is one of the most general organizing principles in physics. By identifying the transformations under which a system and its governing operators remain invariant, symmetry classifies physical states, determines conserved quantities and protected degeneracies, and establishes which couplings and responses are allowed or forbidden. It thereby connects atomic structure to a broad range of electronic, magnetic, optical, mechanical and vibrational properties \cite{wigner2012group, Dresselhaus2008GroupTheory, BradleyCracknell2009}. The same principles classify phonon states and constrain their scattering and transmission \cite{PhysRevB.103.184302, Cen2026}. In insulating and semiconducting solids, where heat is predominantly carried by lattice vibrations, symmetry therefore provides a natural framework for relating the microscopic structure of a material to its thermal-transport properties.

Considering symmetry in quasi-one-dimensional (quasi-1D) crystals such as nanotubes and nanowires requires some care. Conventional atomistic workflows usually represent a quasi-1D crystal as a vacuum-padded three-dimensional periodic crystal and identify its symmetry using an ordinary space-group analysis. Because three-dimensional space groups are subject to the crystallographic restriction theorem, this procedure generally returns only a subgroup of the actual operations of the quasi-1D structure: axial rotations of arbitrary order and generalized translations such as screw operations may be lost. Consequently, calculation workflows that use that information neglect some of the invariances and equivariances. For instance, the interatomic force constants (IFCs) and dynamical matrix used in lattice-dynamical calculations are not guaranteed to fulfill all the symmetries, and phonon modes that should be symmetry-equivalent or degenerate may be artificially separated. In practice, those missing constraints can produce spurious imaginary frequencies, an incorrect number of acoustic modes, errors in sound velocities, splitting or mixing of phonon branches, and finite group velocities in ZA branches that should be quadratic \cite{Carrete2019, Lin2022}. The full symmetry of a quasi-1D crystal is instead described by a line group, whose generalized translations and axial point group restore the correct symmetry relations. Its irreducible representations (irreps), labeled by quantum numbers such as the axial wave vector $k$, angular quasimomentum $m$, and parity $\Pi$, separate the phonon spectrum into well-defined symmetry sectors and provide the constraints needed to calculate and classify the vibrational modes consistently \cite{damnjanovic2010line, Cen2026}.

The accurate calculation and symmetry classification of phonon modes is also the starting point for most detailed descriptions of thermal transport at the atomistic level. Both phonon Boltzmann transport equation (BTE) methods \cite{McGaughey2019, Lindsay2016} and mode-resolved atomistic Green's function (AGF) methods \cite{OngZhang2015,Ong2018} start from harmonic phonon eigenmodes: the BTE describes heat transport through their group velocities and transition rates, whereas AGF approaches resolve transmission from incident to outgoing propagating modes. Assigning line-group irreps to these modes organizes the transport quantities into symmetry sectors and makes the corresponding constraints explicit. In the BTE they restrict anharmonic scattering matrix elements \cite{Lindsay2009,PhysRevB.103.184302}, while in mode-resolved AGF calculations they restrict transmission between incident and outgoing channels \cite{Cen2026}. Nevertheless, these approaches retain an eigenmode-based description, and the associated IFC calculations scale poorly with system size and order of the anharmonicity. 

Molecular dynamics (MD) provides a complementary finite-temperature description of lattice dynamics that samples the full interatomic potential and thereby incorporates the combined effects of anharmonic interactions without imposing a truncation order. Methods such as Green–Kubo modal analysis (GKMA)\cite{Lv_2016,10.1063/1.5081722} and homogeneous nonequilibrium modal analysis (HNEMA)\cite{EVANS1982457,PhysRevB.103.205421} enable a detailed decomposition of the thermal conductivity obtained from MD trajectories. In their conventional formulations, both methods project the atomic velocities onto a phonon eigenbasis. GKMA operates on equilibrium trajectories and evaluates all pairwise correlations between modal heat currents, yielding a complete mode-pair conductivity matrix. HNEMA instead applies a weak homogeneous driving field and obtains a single-index set of modal conductivity contributions within the linear-response regime. Although the individual modal contributions are basis dependent, any complete orthonormal basis yields the same total thermal conductivity\cite{10.1063/1.4921108}. Consequently, if the objective is to resolve the conductivity by irreps rather than by individual phonon branches, it is sufficient to use mutually orthogonal bases spanning the corresponding symmetry-adapted subspaces. Such a basis can be obtained by first block-diagonalizing the dynamical matrix according to symmetry and then diagonalizing each block, yielding a symmetry-adapted phonon eigenbasis (SA eigenbasis) with well-defined line-group labels.

Alternatively, a projector-generated symmetry-adapted basis (projector SAB) can be constructed directly for those subspaces \cite{wigner2012group, M_Damnjanovic_1994, DAMNJANOVIC20151}. That basis can be used in GKMA and HNEMA to obtain irrep-resolved conductivity contributions, without constructing or diagonalizing the dynamical matrix. The projector SAB and SA eigenbasis approaches differ only by a unitary rotation within the same invariant subspaces and therefore yield identical conductivities when summed over a complete irrep block. The approach can be extended from line groups to space groups in three-dimensional systems when the corresponding representations and characters are available.

In this work, we implement a symmetry-adapted framework for both GKMA and HNEMA and apply it to a pristine $(10,0)$--$(20,0)$ \ce{WS2}--\ce{MoS2} double-walled nanotube (DWNT). We compare three different bases: the conventional phonon eigenbasis, the SA eigenbasis obtained by diagonalizing the dynamical matrix within individual irrep blocks, and the projector SAB constructed directly from the symmetry projectors. All three recover the same total thermal conductivity, but the two symmetry-adapted constructions additionally yield identical conductivities when summed within each irrep block. HNEMA resolves how conductivity is distributed among symmetry channels and axial wave vectors, whereas GKMA additionally reveals the correlations between irrep-block currents. Section~\ref{sec:methods} presents the symmetry-adapted GKMA and HNEMA frameworks and the computational methodology. Section~\ref{sec:results} validates the basis invariance and reports the irrep-resolved thermal transport of the DWNT, and Section~\ref{sec:conclusion} summarizes the main findings.

\section{Methods}
\label{sec:methods}

\subsection{GKMA}
\label{subsec:GKMA}

In this framework,\cite{Lv_2016,10.1063/1.5081722} each contribution to the thermal conductivity tensor is written as the time integral of a modal heat-current autocorrelation function,

\begin{equation}
    \kappa^{\alpha\beta}_{\lambda\lambda'} = \frac{1}{\Omega_{\mathrm{eff}}\,k_{\mathrm{B}} T^2}\int_0^\infty \left\langle Q^\alpha_\lambda(0)\, Q^\beta_{\lambda'}(t)\right\rangle \mathrm{d}t
    \label{eqn:modal-kappa}
\end{equation}
where $\Omega_{\mathrm{eff}}$ is the effective volume of the system, $T$ is the temperature, $k_{\mathrm{B}}$ is the Boltzmann constant, $\alpha,\beta$ are Cartesian indices, and $\lambda,\lambda'$ are mode indices.

The modal heat current is obtained by projecting the atomic velocities onto a
modal basis $\{\mathbf{e}_\lambda\}$. The modal velocity coordinate is

\begin{equation}
v_\lambda(t) = \sum_i \sqrt{m_i}\,\mathbf{e}^{*}_{i,\lambda}\cdot\mathbf{v}_i(t),
\end{equation}
where $\mathbf{e}_{i,\lambda}$ denotes the three-component block of $\mathbf{e}_\lambda$ on atom $i$, and the contribution of mode $\lambda$ to the velocity of atom $i$ is $\mathbf{v}_{i,\lambda}(t)=\mathbf{e}_{i,\lambda}\,v_\lambda(t)/\sqrt{m_i}$. The modal heat current then reads

\begin{equation}
Q^{\alpha}_{\lambda}(t) = \sum_i E_i\, v^{\alpha}_{i,\lambda}(t)
+ \sum_i \sum_{j\neq i}
\left( -\frac{\partial \Phi_j}{\partial \mathbf{r}_i} \cdot \mathbf{v}_{i,\lambda}(t) \right)
r^{\alpha}_{ij},
\label{eqn:modal_HC}
\end{equation}
While the kinetic contribution to the total energy $E_i=\tfrac12 m_i|\mathbf v_i|^2+\Phi_i$ is unambiguous, the interaction energy $\Phi_i$ may be redistributed among atoms without changing the total energy. Modern machine learning interatomic potentials, including the neuroevolution potential (NEP) used in this paper, are actually well suited for this task, as they model the total system energy as a sum of atomic terms, even though they do not have specific physical meaning. These atomic pseudoenergies are not themselves unique observables: different partitions of the same total energy define different local-energy gauges and can yield different instantaneous microscopic heat currents. Nevertheless, the Green--Kubo thermal conductivity obtained from the full heat current is invariant under such gauge transformations~\cite{Ercole2016}. 

In conventional GKMA the modal basis $\left\{\bm{e}_{\lambda}\right\}$ is taken to be the set of eigenvectors of the harmonic dynamical matrix of the entire simulation supercell at the $\Gamma$ point. For a supercell containing $K$ primitive cells, this $\Gamma$-point displacement space is unitarily equivalent to the direct sum of the primitive-cell phonon spaces at the $K$ commensurate wave vectors $k_n=2\pi n/\left(Ka\right)$ that fold onto the supercell $\Gamma$ point. Resolving the supercell displacement space according to primitive translations therefore recovers the $k_n$ labels used below.

\subsection{HNEMA}
\label{subsec:HNEMA}

HNEMA\cite{PhysRevB.103.205421} decomposes the lattice thermal conductivity into per-mode contributions in a homogeneous nonequilibrium MD framework (HNEMD)\cite{EVANS1982457}. In HNEMD, a small homogeneous driving field $\bm{F}_{\mathrm{e}}$ is imposed along the transport direction, generating a steady-state heat current $\aqty*{J^{\alpha}_{\lambda}(t)}_{\mathrm{ne}}$ in the linear-response regime. The mode-resolved thermal conductivity is obtained from

\begin{equation}
    \frac{\aqty*{J^{\alpha}_{\lambda}(t)}_{\mathrm{ne}}}{T\Omega_{\mathrm{eff}}} = \sum_{\beta}\kappa^{\alpha\beta}_{\lambda}F_{\mathrm{e}}^{\beta}
\end{equation}
where $\kappa_{\lambda}^{\alpha\beta}$ is the thermal conductivity tensor of mode $\lambda$. The modal heat current $J^{\alpha}_{\lambda}(t)$ is constructed in exactly the same way as as in Eq.~\ref{eqn:modal_HC} for GKMA.

In contrast to the per-mode-pair structure of Eq.~\eqref{eqn:modal-kappa}, HNEMA yields a single-mode-index modal decomposition of thermal conductivity. Within the linear-response framework, the external driving field $\bm{F}_{\mathrm{e}}$ induces a nonequilibrium heat current, and the resulting modal contribution is obtained from the projected modal heat current $\aqty*{J^{\alpha}_{\lambda}(t)}_{\mathrm{ne}}$. Thus, HNEMA assigns a net conductivity contribution to each mode $\lambda$, but it does not explicitly resolve mode-mode cross correlations or a full $\kappa_{\lambda \lambda'}$ matrix.
HNEMA is therefore complementary to GKMA, providing a statistically efficient route to single-mode conductivities that converges one to two orders of magnitude faster than the equilibrium Green--Kubo methods \cite{PhysRevB.103.205421} while GKMA remains a natural tool for analyzing cross-mode contributions through Eq.~\eqref{eqn:modal-kappa}. In the following sections we develop a symmetry-adapted construction of modal basis $\{\bm{e}_{\lambda}\}$ that applies to both methods.

\subsection{Line-group symmetry and projectors for quasi-1D systems}
\label{subsec:SAB}

A line group $L=Z\cdot P$ is the semidirect product of a generalized translational group $Z$, which may take the form of a pure translation, a screw axis $\pqty*{C_{Q} \mid f}$ or a glide plane $\pqty*{\sigma_{V} \mid a/2}$ and an axial point group $P$ drawn from one of seven families ($C_{n}$, $S_{2n}$, $C_{nh}$, $D_{n}$, $C_{nv}$, $D_{nd}$, $D_{nh}$) \cite{damnjanovic2010line}. Here $C_{Q}$ denotes a $Q$-fold rotation about the axial direction, $\sigma_{V}$ is a vertical mirror reflection, $f$ is the translation along the axis accompanying the screw rotation, and $a$ is the primitive axial translation. In the family labels, $n$ denotes the order of the principal rotation axis. Each irrep $\mu$ of the line group is parameterized by quantum numbers: the axial wave vector $k$, the angular-momentum index $m$ and parity labels $\Pi_{\mathrm{V}}$, $\Pi_{\mathrm{H}}$ or $\Pi_{\mathrm{U}}$.
Using the package \textsc{Pulgon-tools}~\cite{cen2026pulgontoolstoolkitanalysingharnessing}, we identify the line group of input structures and obtain the character $\chi^{\mu}(g)$ for each line group element $g$ in each irrep $\mu$.

Under the periodic boundary conditions of the finite simulation supercell, the generalized translation subgroup is taken modulo the supercell period. The axial wave vector $k$ is consequently restricted to the commensurate set $k_{n}=2\pi n/\left(Ka\right)$, with $n=0, \ldots, K-1$. This construction gives a finite supercell factor group $G$, for which we construct the symmetry projector at allowed wave vector $k_{n}$ and irrep label $\mu$ \cite{wigner2012group,M_Damnjanovic_1994,DAMNJANOVIC20151}:
\begin{equation}
    \bm{P}^{\mu}(k) = \frac{d_{\mu}}{|G|}\sum_{g}\chi^{\mu}(g)^{*}\bm{S}(g)
    \label{eqn:projector}
\end{equation}
where $d_{\mu}$ is the dimension of the irrep and $|G|$ is the order of this finite supercell group. Because Eq.~\eqref{eqn:projector} is a character projector, it fixes the full irrep block, including all repeated copies of $\mu$, but does not select a unique basis within its multiplicity space. The irrep-block sums reported below are invariant under unitary rotations in this space. The matrix $\bm{S}(g)$ acts on the $3N$-dimensional space of atomic displacements in the simulation box, and is defined by the displacement transformation matrix $\bm{M}(g)$ and two phase factors:

\begin{equation}
    \bm{S}_{ij}^{\alpha\beta}(g) = \bm{M}_{ij}^{\alpha\beta}(g) \cdot e^{ikT_{z}} \cdot e^{ik \left(r_{i}^{z} - r_{j}^{z}\right)}
\label{eqn:rep_max}
\end{equation}
where $\bm{M}(g)$ is constructed by placing copies of the three-dimensional Cartesian rotation matrix associated with $g$ at the rows and columns defined by the atom permutation induced by the symmetry operation. The first phase factor $e^{ikT_{z}}$ is associated with the translational component of $g$, where $T_{z}$ is the component of its translational vector along the periodic axis. The second exponential factor $e^{ik \left(r_{i}^{z} - r_{j}^{z}\right)}$ accounts for the Bloch phase difference between the source and target atoms.

For each projector we extract its orthonormal basis, $\bm{B}^{\mu}(k)=\bqty*{\bm{b}^{\mu k}_{1},\ldots,\bm{b}^{\mu k}_{r_{\mu k}}}$, where $r_{\mu k}=\mathrm{rank}\,\bm{P}^{\mu}(k)$. Concatenating these bases across all irreps $\mu$ at a single wave vector $k$ yields the projector SAB at that $k$, which block-diagonalizes the dynamical matrix $\bm{D}(k)$ in the primitive cell.

For the modal decomposition in GKMA and HNEMA, this construction is applied at each allowed folded wave vector. For every pair $\left(k_n,\mu\right)$ we construct $\bm{P}^{\mu}\left(k_n\right)$, extract its column-space basis $\bm{B}^{\mu n}$, and only then concatenate the resulting bases. Thus the projectors for different $k_n$ are not first combined into a single operator. The complete projector SAB of the supercell is

\begin{equation}
    \ket{\mathrm{SAB}} =
    \bqty{\bm{B}^{\mu_1 0},\ldots,\bm{B}^{\mu_M 0},
    \bm{B}^{\mu_1 1},\ldots,\bm{B}^{\mu_M,K-1}},
    \label{eqn:orthS}
\end{equation}
where $\mu_1,\ldots,\mu_M$ denote the irreps present at a given $k_n$. Each column of the projector SAB is a basis vector of the supercell displacement space carrying definite line-group quantum numbers $\pqty*{k_n,m,\Pi_{\mathrm{V}}}$. We denote the $\lambda$-th column by $\ket{\mathrm{SAB}}_{\lambda}$.

\subsection{Symmetry-adapted modal bases}
\label{subsec:sym-eig}

The modal analysis frameworks introduced in Sec.~\ref{subsec:GKMA} and \ref{subsec:HNEMA} are conventionally implemented using phonon eigenvectors obtained by diagonalizing the supercell dynamical matrix at $\Gamma$, for example with \textsc{Phonopy} \cite{phonopy-phono3py-JPCM,phonopy-phono3py-JPSJ}. We now consider two modal bases adapted to the line-group symmetry of the supercell.

Starting from the projector SAB of Eq.~\eqref{eqn:orthS}, we consider two symmetry-adapted constructions of the modal basis $\bm{e}_\lambda$. The first is to use the projector SAB directly. It is determined entirely by the line group and the supercell geometry, independent of the dynamical matrix or the interatomic potential. The irrep-block decomposition into subspaces is unique, but the individual basis vectors are not: any unitary rotation within a given irrep block yields an equally valid SAB. The second is to first transform the supercell $\Gamma$-point dynamical matrix into the projector SAB. The blocks of the transformed matrix are labeled by the folded-wave-vector and irrep labels $(k_n,\mu)$; each block is then diagonalize each block separately. This yields the SA eigenbasis $\{\bm{e}^{\mathrm{SA}}_\lambda\}$, whose vectors represent phonon modes carrying line-group quantum-number labels. The dynamical-matrix diagonalization is therefore performed after the $(k_n,\mu)$ projector bases have been concatenated, and only within the symmetry blocks.

In both constructions, the modal basis is complete and orthonormal, so the modal heat currents Eq.~\eqref{eqn:modal_HC} add up to the total heat current and the total thermal conductivity is preserved.

Replacing the eigenvector basis in Eq.~\eqref{eqn:modal-kappa} by either of the symmetry-adapted constructions yields an irrep-resolved thermal conductivity matrix in GKMA:

\begin{equation}
        \kappa^{\alpha\beta}_{\mu,\nu} = \frac{1}{\Omega_{\mathrm{eff}}\,k_{\mathrm{B}}T^{2}}\int^{\infty}_{0}\aqty*{Q^{\alpha}_{\mu}(0)Q^{\beta}_{\nu}(t)}\mathrm{d}t
    \label{eqn:irrep-kappa}
\end{equation}
where $Q^{\alpha}_{\mu}=\sum_{\lambda\in\mu}Q^{\alpha}_{\lambda}$ is the modal heat current summed over all modes in irrep block $\mu$, labeled by the complete tuple $(k,m,\Pi_{\mathrm{V}})$. The diagonal terms $\kappa_{\mu\mu}^{\alpha\beta}$ are the self-correlation contributions of individual irrep blocks, while the off-diagonal entries $\kappa_{\mu\nu}^{\alpha\beta}$ ($\mu \neq \nu$) encode correlations between distinct blocks. Sums over $k$ are referred to below as symmetry-channel contributions or symmetry-channel-pair matrices.

There are two advantages of this construction compared with conventional GKMA. Firstly, each vector in the projector SAB belongs to a definite $\left(k_n,\mu\right)$ irrep subspace and therefore carries line-group quantum numbers regardless of dynamical-matrix degeneracy. Secondly, the projector SAB itself requires no diagonalization of the harmonic dynamical matrix, since it is determined entirely by the symmetry of the structure; if phonon eigenvectors are desired, the subsequent diagonalization is confined to the symmetry-adapted blocks of the dynamical matrix.

\subsection{Training of the interatomic potential}
\label{subsec:GPUMD}

All training labels were obtained from single-point density-functional-theory (DFT) calculations. We use the projector augmented-wave formalism~\cite{PhysRevB.50.17953} as implemented in the 6.5.1 release of VASP~\cite{PhysRevB.54.11169, KRESSE199615, PhysRevB.59.1758}, together with the PBEsol approximation to the exchange and correlation energy~\cite{PhysRevLett.100.136406} and the DFT-D3 method with a Becke--Johnson damping function~\cite{10.1002/jcc.21759} to approximately account for van der Waals interactions. We choose $5s^{1}4d^{5}$, $6s^{2}5d^{4}$ and $3s^{2}3p^{4}$ as valence configurations for Mo, W and S respectively, a plane-wave cutoff energy of \SI{600}{\eV}, and $1\times 1\times 2$, $1\times 1\times 2$ and $2\times 2\times 1$ $\Gamma$-centred Monkhorst--Pack $k$-point grids for single-walled nanotubes, DWNT and 2D structures respectively. We moreover set an electronic convergence criterion of \SI{e-6}{\eV} and a Gaussian smearing of \SI{0.01}{\eV} for all single-point runs. For each configuration the total energy, atomic forces, and stress tensor were used as regression targets.

The interatomic potential is a locally descriptor-based NEP natively implemented in GPUMD \cite{https://doi.org/10.1002/mgea.70028}. NEP is well suited to the present work: it runs efficiently on GPUs, the heat-current decomposition required by GKMA \cite{Lv_2016} and HNEMA \cite{PhysRevB.103.205421} is built in, and the modal basis used in that decomposition can be supplied externally, which is essential for the symmetry-adapted constructions of Sec.~\ref{subsec:sym-eig}.

The training configurations sample the WS$_2$ and MoS$_2$ monolayers, the pristine WS$_2$ $(10,0)$ and MoS$_2$ $(20,0)$ single-wall nanotubes, and the $(10,0)$--$(20,0)$ WS$_2$--MoS$_2$ DWNT. Finite-temperature snapshots at $\SI{300}{K}$ were drawn both from short ab-initio MD trajectories and from MD using the MACE-MATPES-PBE-0 foundation model~\cite{10.1063/5.0297006, kaplan2025foundationalpotentialenergysurface}, and were augmented by elastically deformed structures to sample the strained regime. After single-point DFT labeling this yields a dataset of $920$ configurations, partitioned into $820$ training and $100$ test structures. A single production potential was then trained on this dataset for $10^{5}$ generations of the separable natural-evolution strategy, using cutoffs of $\SI{7}{\angstrom}$ and $\SI{5}{\angstrom}$ for the radial and angular descriptors respectively and including energy, force, and virial in the loss function. The resulting energy and force accuracies are reported in Appendix~\ref{subsec:NEP-accuracy} (Fig.~\ref{fig:RMSE}).

All MD trajectories were generated with this potential in GPUMD~4.2 \cite{https://doi.org/10.1002/mgea.70028}, using a timestep of $\SI{1}{fs}$. Each trajectory was initialized with velocities drawn from a Maxwell--Boltzmann distribution at the target temperature and equilibrated for $\SI{100}{ps}$ in the canonical (NVT) ensemble. The HNEMA and HNEMD production runs remained in the NVT ensemble, with the temperature controlled by a Nos\'e--Hoover chain thermostat with a coupling time of $\SI{100}{fs}$. For GKMA, the thermostat was removed after equilibration and the production trajectories were generated in the microcanonical (NVE) ensemble.

The application system is the pristine $(10,0)$--$(20,0)$ WS$_2$--MoS$_2$ DWNT, whose line group is $T'C_{10}$ (family~7) with a translational period of $180$ atoms ($40$~Mo, $20$~W, $120$~S). All transport calculations use a $[1,1,10]$ simulation supercell, i.e.\ $K=10$ primitive cells along the tube axis. The allowed axial wave vectors are therefore $k_n = 2\pi n/(10a)$ with $n=0,\dots,9$, where $a$ is the primitive-cell length. Following the family-7 convention of Table~4.7 in Ref.~\cite{damnjanovic2010line}, the eight symmetry-channel labels at each $k$ are denoted $m0A=(0,-1)$, $m0B=(0,+1)$, $m1E=(1,0)$, $m2E=(2,0)$, $m3E=(3,0)$, $m4E=(4,0)$, $m5A=(5,-1)$, and $m5B=(5,+1)$, where each pair gives $(m,\Pi_{\mathrm{V}})$. Thus $A$ and $B$ denote $\Pi_{\mathrm{V}}=-1$ and $+1$, respectively, while $E$ denotes the doubly degenerate channels with $m=1$--$4$. Here $V$ refers to the glide plane in the generalized translational group $T'$. Its one-dimensional characters contain the factor $\Pi_{\mathrm{V}}e^{ika/2}$. Combined with the ten allowed wave vectors, these channels yield $8\times 10 = 80$ irrep blocks that label the projector SAB of the supercell.

Modal conductivities are reported per unit effective volume $\Omega_{\mathrm{eff}}$ [Eq.~\eqref{eqn:modal-kappa}]. We define $\Omega_{\mathrm{eff}}=\pi(R_{\mathrm{out}}^2-R_{\mathrm{in}}^2)L$, where $R_{\mathrm{in}}$ and $R_{\mathrm{out}}$ are the minimum and maximum atomic-center distances from the tube axis. The annulus is therefore bounded by the radial positions of the innermost and outermost nuclei, with no additional wall-thickness parameter. For the $[1,1,10]$ simulation supercell, $R_{\mathrm{in}}=\SI{3.925}{\angstrom}$, $R_{\mathrm{out}}=\SI{12.235}{\angstrom}$, and $L=\SI{55.457}{\angstrom}$, giving $\Omega_{\mathrm{eff}}=\SI{2.3398e4}{\cubic\angstrom}$. This choice rescales all modal conductivities by a common factor and therefore leaves all conductivity fractions and channel-partition ratios unaffected.

For the choice of driving field $F^{\mathrm{drv}}$ in HNEMA, the linear-response regime was checked using homogeneous driving fields of $\SI{2e-6}{\per\angstrom}$, $\SI{4e-6}{\per\angstrom}$, $\SI{6e-6}{\per\angstrom}$, and $\SI{1e-5}{\per\angstrom}$ (Appendix~\ref{app:suppfigs}, Fig.~\ref{fig:hnemd_fe_scan}).
After testing a homogeneous driving field of magnitude $\SI{4e-6}{\per\angstrom}$ is imposed along the tube axis for the main HNEMA result. This is a common field parameter, not an identical force on every atom. In GPUMD the atom-dependent driving force is formed from the per-atom virial tensor $W_i$ as $F^{\mathrm{drv}}_{i\alpha}=\sum_{\beta}F_{{\mathrm{e}},\beta}W_{i,\beta\alpha}$, after which the system-mean force is subtracted to maintain zero net force. Eight independent $\SI{100}{ns}$ trajectories are used for the main result. Each trajectory provides 500 consecutive interval modal-conductivity estimates, one per $\SI{0.2}{ns}$ interval, and all estimates from $0$ to $\SI{100}{ns}$ are averaged to obtain one seed-resolved value. The reported conductivity is the mean over the eight seed-resolved full-trajectory estimators, and its uncertainty is their standard error.

For GKMA, six independent production trajectories are generated for $\SI{100}{\nano\second}$, with modal heat currents sampled every $\SI{10}{fs}$. The cutoff scan in Fig.~\ref{fig:dwnt_gkma_conv} identifies the first broad plateau of the six-seed mean from $\SI{0.30}{ns}$ to $\SI{0.70}{ns}$. For each complete trajectory, the irrep-resolved conductivity matrix is evaluated at nine correlation-time cutoffs in this interval, spaced by $\SI{0.05}{ns}$, and averaged over these cutoffs to obtain one seed-resolved matrix. The reported matrix is the average over the six seeds. The total conductivity and its uncertainty are the mean and standard error of the six seed-resolved totals. The block-length test is also shown in Fig.~\ref{fig:dwnt_gkma_conv}.

\section{Results and discussion}
\label{sec:results}

\subsection{Structure and symmetry-resolved phonons}
\label{subsec:structure-phonons}

\begin{figure}[tbp]
    \centering
    \includegraphics[width=1\linewidth]{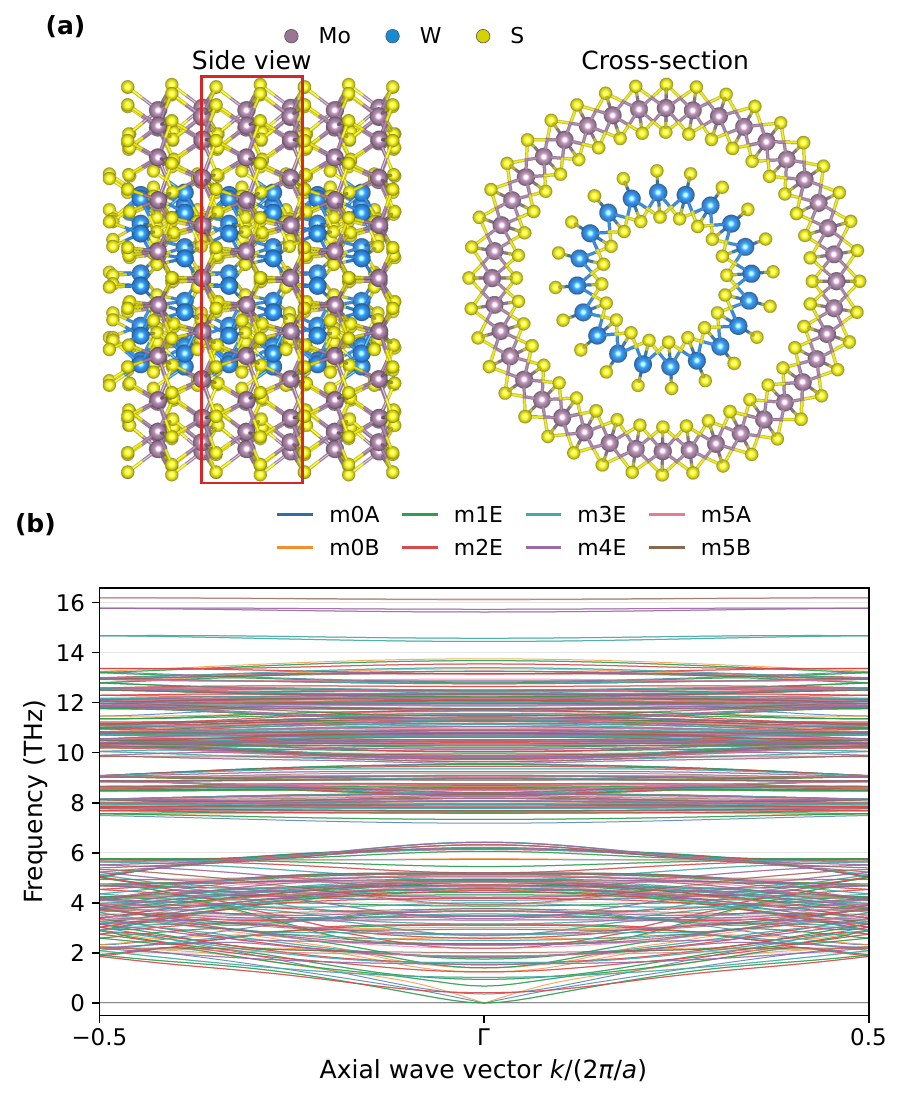}
    \caption{Atomic structure and symmetry-resolved harmonic phonons of the pristine $(10,0)$--$(20,0)$ WS$_2$--MoS$_2$ DWNT. (a) Side and cross-sectional views, with W, Mo, and S atoms shown in blue, purple, and yellow, respectively. The red rectangle marks the primitive cell. (b) Harmonic phonon spectrum, colored according to the eight line-group symmetry channels. For the one-dimensional channels with $m=0$ and $5$, $A$ and $B$ denote $\Pi_{\mathrm{V}}=-1$ and $+1$, respectively. $m=1$--$4$ are the doubly degenerate $E$-type channels.}
    \label{fig:structure-phonons}
\end{figure}

The pristine DWNT consists of a $(10,0)$ WS$_2$ inner tube coaxially enclosed by a $(20,0)$ MoS$_2$ outer tube [Fig.~\ref{fig:structure-phonons}(a)]. The two walls have a common axial period and preserve the tenfold line-group symmetry of the combined structure. The primitive cell contains 180 atoms (20 \ce{W}, 40 \ce{Mo} and 120 \ce{S}) and is used to construct the $[1,1,10]$ transport supercell described in Sec.~\ref{subsec:GPUMD}.

The symmetry-resolved phonon spectrum in Fig.~\ref{fig:structure-phonons}(b) contains the expected $3N=540$ phonon branches, including four acoustic branches. Each branch is assigned to a definite $(m,\Pi_{\mathrm{V}})$ symmetry channel throughout the one-dimensional Brillouin zone. Sampling this spectrum at the ten wave vectors allowed by the $[1,1,10]$ supercell gives the 80 $(k,m,\Pi_{\mathrm{V}})$ irrep blocks used below to resolve the HNEMA and GKMA conductivities.

The irrep labels describe how an atomic displacement pattern transforms: $k$ specifies its phase change under axial translation, $m$ its behavior under axial rotation, and $\Pi_{\mathrm{V}}$ its glide parity where defined. An irrep block can contain multiple phonon modes with different frequencies and displacement patterns. Representative examples are shown in Appendix~\ref{app:irrep-patterns}.

\subsection{Comparison of the three modal bases}
\label{subsec:sym-eig-compare}

To examine the differences among the conventional phonon eigenbasis, the SA eigenbasis and the projector SAB, we compare their diagonal modal HNEMA thermal conductivities for the WS$_2$--MoS$_2$ DWNT. All three bases are evaluated on the same eight independent HNEMA runs over a common 10--$\SI{40}{ns}$ diagnostic window, so that they differ only in the modal vectors and not in the underlying dynamics.

\begin{figure}[tbp]
    \centering
    \includegraphics[width=1\linewidth]{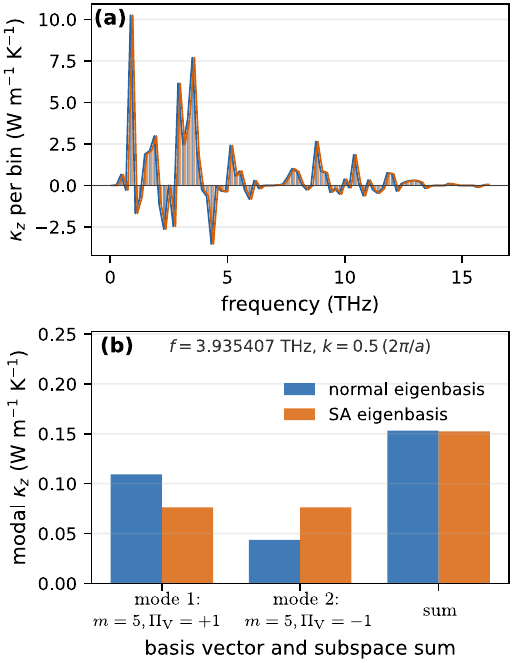}
    \caption{Comparison of the HNEMA modal axial thermal conductivity $\kappa_z$ obtained with the normal eigenbasis and the SA eigenbasis. (a) Frequency-binned conductivity over the full spectrum. (b) A selected doubly degenerate subspace at $f=\SI{3.935407}{THz}$ and $k=0.5\,(2\pi/a)$. The SA eigenbasis bars are the parity-resolved modes $(m,\Pi_{\mathrm{V}})=(5,+1)$ and $(5,-1)$. The final bars show the invariant sum over the doublet. }
    \label{fig:hnema_normal_vs_sym}
\end{figure}

Figure~\ref{fig:hnema_normal_vs_sym}(a) shows the axial thermal conductivity in each frequency bin for the conventional phonon eigenbasis and the SA eigenbasis. The two distributions are in excellent agreement across the full frequency range, confirming that the frequency-resolved conductivity is unchanged by adapting the eigenbasis to the line-group symmetry.
The difference between the two eigenbases becomes apparent only at the level of individual degenerate modes. Figure~\ref{fig:hnema_normal_vs_sym}(b) shows a zone-boundary doublet whose two SA eigenbasis modes have the same frequency and wave vector but opposite parities, $m5B$ and $m5A$. Direct projection identifies the corresponding conventional phonon doublet as spanning the same two-dimensional subspace. Transforming this subspace into the parity-resolved SA eigenbasis redistributes the two modal conductivities from $0.109$ and $\SI{0.044}{\watt\per\meter\per\kelvin}$ to $0.076$ and $\SI{0.076}{\watt\per\meter\per\kelvin}$, putting them on equal footing while their sum is unchanged.

\begin{figure}[tbp]
    \centering
    \includegraphics[width=1\linewidth]{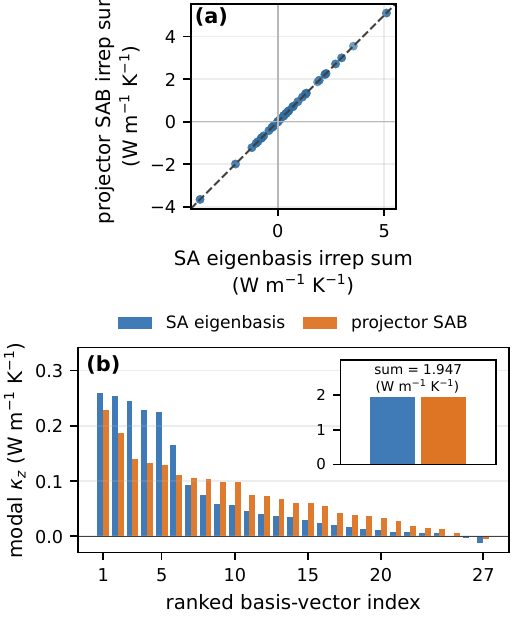}
    \caption{Comparison of HNEMA diagonal modal axial thermal conductivity $\kappa_z$ of the WS$_2$--MoS$_2$ DWNT obtained using the SA eigenbasis and the projector SAB. (a) Conductivity within an irrep block. Each of the 80 points corresponds to one complete $\pqty*{k,m,\Pi_{\mathrm{V}}}$ label and the dashed line indicates perfect agreement between the two bases. (b) Ranked mode-resolved contributions within the 27-dimensional irrep block $\pqty*{k,m,\Pi_{\mathrm{V}}}=(0.4,5,-1)$. The inset compares the summed conductivity of the complete block.}
    \label{fig:hnema_sym_vs_projector}
\end{figure}

We further compare the SA eigenbasis with the projector SAB in Fig.~\ref{fig:hnema_sym_vs_projector}. Figure~\ref{fig:hnema_sym_vs_projector}(a) presents the thermal conductivity summed over all modes in each of the 80 irrep blocks that span the 5400 modes of the DWNT. The data points lie on the diagonal line, indicating that every irrep-block sum is identical for the two bases. This demonstrates that the physically meaningful conductivity associated with a complete $(k,m,\Pi_{\mathrm{V}})$ label is invariant under different basis constructions inside the same symmetry subspace. The SA eigenbasis and projector SAB therefore span identical irrep blocks and differ only by a unitary rotation within each block.

Figure~\ref{fig:hnema_sym_vs_projector}(b) illustrates the resulting mode-level basis dependence within the irrep block $ \pqty*{k,m,\Pi_{\mathrm{V}}}=(0.4,5,-1)$. Its summed conductivity is $\SI{1.947}{\watt\per\meter\per\kelvin}$ in both bases, whereas the individual contributions are redistributed. The projector SAB is constructed solely from the line-group symmetry and supercell geometry and therefore contains no material-specific dynamical information. It yields the same summed conductivity for each irrep block as the SA eigenbasis, but its individual vectors are not associated with unique phonon modes. The comparison shows that the modal contributions are redistributed among individual modes while the sum is preserved.

\subsection{HNEMA transport in the DWNT}
\label{subsec:Tdep}

\begin{figure}[htbp]
    \centering
    \figph{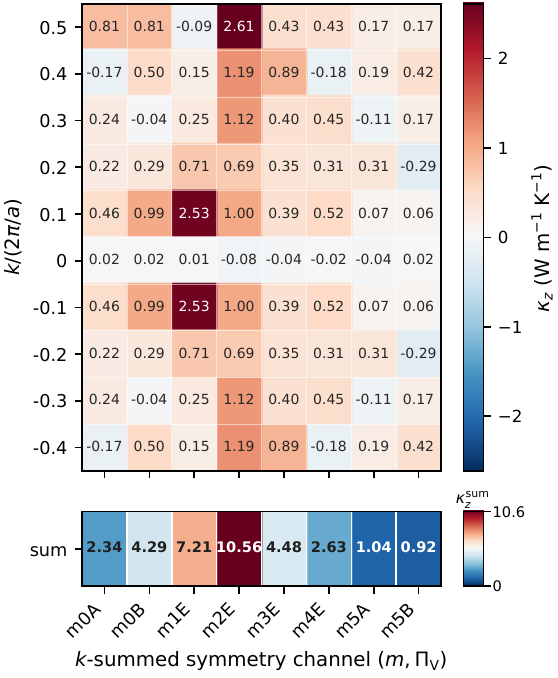}
    \caption{$\pqty*{k,m,\Pi_{\mathrm{V}}}$-resolved HNEMA thermal conductivity of the DWNT. The main grid contains the contributions of the 80 irrep blocks resolved by the ten axial wave vectors $k$ (in units of $2\pi/a$) and eight symmetry channels. The bottom row gives the $k$-summed seed mean for each channel. For the one-dimensional channels at $m=0$ and $5$, $A$ and $B$ denote $\Pi_{\mathrm{V}}=-1$ and $+1$, respectively. $m=1$--$4$ are doubly degenerate $E$-type channels. All values are in \si{\watt\per\meter\per\kelvin}.}
    \label{fig:dwnt_hnema_km}
\end{figure}

We now apply HNEMA in the SA eigenbasis to the pristine $(10,0)$--$(20,0)$ WS$_2$--MoS$_2$ DWNT, whose line group is $T'C_{10}$ with eight $\pqty*{m,\Pi_{\mathrm{V}}}$ symmetry channels at each $k$. Using the $[1,1,10]$ supercell and the driven steady-state protocol of Sec.~\ref{subsec:GPUMD}, the eight seed-resolved full-trajectory estimators give a mean total conductivity of $\SI{33.5}{\watt\per\meter\per\kelvin}$ at $\SI{300}{K}$, with a seed standard error of $\SI{3.6}{\watt\per\meter\per\kelvin}$ as shown in Fig.~\ref{fig:dwnt_hnema_conv300}. HNEMA yields one conductivity contribution for each modal basis vector and does not construct a mode-pair matrix. Figure~\ref{fig:dwnt_hnema_km} therefore displays the single-index contributions summed within each $(k,m,\Pi_{\mathrm{V}})$ irrep block, together with the corresponding $k$-summed symmetry channels.

The two largest $k$-summed mean contributions are those of $m2E$ and $m1E$, $\SI{10.56\pm1.13}{\watt\per\meter\per\kelvin}$ and $\SI{7.21\pm3.95}{\watt\per\meter\per\kelvin}$, respectively. The four $E$-type channels together contribute $\SI{24.88}{\watt\per\meter\per\kelvin}$, $74.3\%$ of the total, with a jackknife standard error of $4.4 \%$. This is consistent with its proportion of modes: the $E$-type account for 4320 of the 5400 modes ($80\%$). The two $m=5$ channels have a combined mean of $\SI{1.96\pm1.2}{\watt\per\meter\per\kelvin}$. The $k$-resolved distribution is concentrated in paired nonzero-wave-vector sectors: the $k=\pm0.1\,(2\pi/a)$ rows together contribute $\SI{12.06\pm3.12}{\watt\per\meter\per\kelvin}$ and the $k=\pm0.4\,(2\pi/a)$ rows contribute $\SI{6.01\pm1.70}{\watt\per\meter\per\kelvin}$, whereas the $\Gamma$-point row is statistically consistent with zero.

\subsection{GKMA analysis in the DWNT}
\label{subsec:cross}

We now turn to the equilibrium GKMA decomposition of the same DWNT, which retains the full mode-pair structure of the conductivity. Using the averaging procedure described in Sec.~\ref{subsec:GPUMD}, we obtain a total axial conductivity of $\SI{33.2\pm2.5}{\watt\per\meter\per\kelvin}$. This agrees with the HNEMA result of $\SI{33.5\pm3.6}{\watt\per\meter\per\kelvin}$ within the uncertainty of either estimate. The convergence of the thermal conductivity of the GKMA with correlation time is shown in Appendix~\ref{app:suppfigs}, Fig.~\ref{fig:dwnt_gkma_conv}.

\begin{figure}[htbp]
    \centering
    \figph{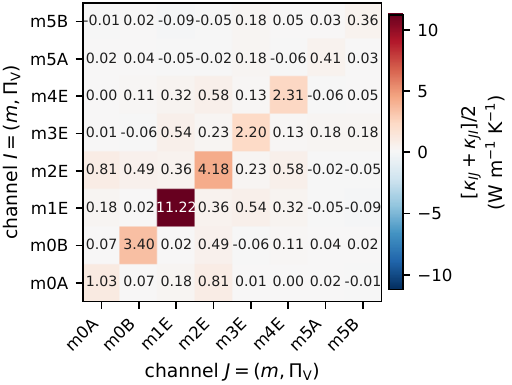}
    \caption{Symmetry-channel-resolved GKMA $8\times8$ $(m,\Pi_{\mathrm{V}})$ thermal-conductivity matrix of the DWNT at $\SI{300}{K}$. Entries are the symmetrized values $\frac{1}{2}(\kappa_{IJ}+\kappa_{JI})$ in \si{\watt\per\meter\per\kelvin}. For the $m=0$ and $5$ channels, $A$ and $B$ denote $\Pi_{\mathrm{V}}=-1$ and $+1$, respectively. $m=1$--$4$ are doubly degenerate $E$-type channels.}
    \label{fig:dwnt_gkma_channels}
\end{figure}

Unlike HNEMA, which yields individual modal contributions with a single irrep tag, GKMA retains the complete irrep-block-pair conductivity matrix. Combining the ten axial wave vectors with the eight $\pqty*{m,\Pi_{\mathrm{V}}}$ channels gives 80 irrep blocks and hence an $80\times80$ matrix. After summing over both wave-vector indices, this matrix reduces to the $8\times8$ symmetry-channel-pair matrix shown in Fig.~\ref{fig:dwnt_gkma_channels}. Its off-diagonal entries quantify the contributions to the total conductivity arising from time-integrated cross correlations between distinct channel-current components. The underlying $80\times80$ matrix is shown in Appendix~\ref{app:suppfigs} (Fig.~\ref{fig:dwnt_gkma_matrix}).

To clarify the symmetry of the pair matrix itself, we note that the modal conductivity of Eq.~\eqref{eqn:irrep-kappa} is a one-sided time integral, $\kappa_{IJ}\propto\int_0^{\infty} C_{IJ}(t)\,dt$, with $C_{IJ}(t)=\aqty*{Q_I(0)Q_J(t)}$. Stationarity gives $C_{IJ}(-t)=C_{JI}(t)$. In addition, every modal heat current in Eq.~\ref{eqn:modal_HC} is odd under time reversal. The zero-field equilibrium ensemble is invariant under this transformation, so $C_{IJ}(t)=C_{IJ}(-t)$. Combining these two relations gives
\begin{equation}
    C_{IJ}(t)=C_{IJ}(-t)=C_{JI}(t),
    \label{eqn:onsager}
\end{equation}
and therefore the exact GKMA matrix obeys the Onsager relation $\kappa_{IJ}=\kappa_{JI}$. A finite trajectory does not satisfy this equality exactly. We therefore average the two finite-sample estimators $\kappa_{IJ}$ and $\kappa_{JI}$, which are equal in the equilibrium limit, thereby using both lag directions and reducing sampling noise. Symmetrization preserves both the total conductivity and the partition sums.

The irrep-block-pair conductivity matrix $\kappa_{IJ}$ is the time-integrated cross correlation of two projected heat-current components. A positive pair contribution means that the corresponding channel-current fluctuations reinforce each other in the total conductivity, whereas a negative contribution represents cancellation.

This pair resolution exposes information that is unavailable in the single-index HNEMA decomposition. At $\SI{300}{K}$, within-block correlations with the same $\pqty*{k,m,\Pi_{\mathrm{V}}}$ label contribute $\SI{40.9\pm2.8}{\%}$ of the total conductivity, while correlations between different $k$ blocks in the same $(m,\Pi_{\mathrm{V}})$ channel contribute $\SI{34.8\pm1.9}{\%}$. Thus $\SI{75.6\pm4.4}{\%}$ of the total retains the same $(m,\Pi_{\mathrm{V}})$ channel label, whereas cross-channel correlations contribute $\SI{24.4\pm4.4}{\%}$. The $k$-resolved matrix (Appendix~\ref{app:suppfigs}, Fig.~\ref{fig:dwnt_gkma_matrix}) further shows that the prominent cross-$k$ contributions primarily connect the momentum-conjugate sectors $k_n$ and $k_{10-n}$. The GKMA matrix therefore resolves how symmetry-labeled current fluctuations combine and cancel to produce the total conductivity. The per-seed stability of this partition is shown in Appendix~\ref{app:suppfigs}, Fig.~\ref{fig:dwnt_gkma_partition}.

\section{Conclusion}
\label{sec:conclusion}

We have extended MD-based modal analysis to incorporate line-group symmetry from the outset through the projector SAB and the SA eigenbasis, in which every modal coordinate carries definite line-group quantum numbers $\pqty*{k,m,\Pi_{\mathrm{V}}}$. With either basis, GKMA and HNEMA yield a unique, irrep-resolved thermal conductivity, free of the arbitrary unitary mixing that afflicts degenerate eigenvectors in conventional implementations. The two methods are complementary: HNEMA provides a statistically efficient single-index decomposition into modal and per-channel contributions, whereas GKMA recovers the full irrep-block-pair matrix, including the off-diagonal cross-channel correlations that HNEMA cannot access. 

Applying the framework to the pristine $(10,0)$--$(20,0)$ WS$_2$--MoS$_2$ DWNT, we show that the SA eigenbasis and projector SAB give identical irrep-block sums, confirming their invariance under basis rotations within each block. Both decompositions reveal that transport is distributed over several symmetry sectors rather than confined to a single channel. HNEMA and GKMA assign the largest mean conductivities to $m2E$ and $m1E$, while HNEMA's $k$-resolved contribution is concentrated in paired nonzero-wave-vector sectors rather than at the $\Gamma$ point. GKMA further resolves how these symmetry-labeled current components are correlated: within-block and same-channel cross-$k$ terms together account for $75.6\%$ of the conductivity at $\SI{300}{K}$, with the prominent cross-$k$ contributions primarily connecting momentum-conjugate sectors, while cross-channel correlations account for the remaining $24.4\%$.

The framework therefore provides a basis-invariant decomposition of MD thermal transport at the irrep-block level, while GKMA additionally reveals how the resulting symmetry-labeled current components correlate in the total conductivity. Because the projector SAB avoids constructing and diagonalizing the dynamical matrix, its computational advantages for larger systems merit further investigation. 

\begin{acknowledgments}
This research was funded in part by the Austrian Science Fund (FWF) [10.55776/P36129]. For open access purposes, the author has applied a CC BY public copyright license to any author-accepted manuscript version arising from this submission. It was also supported by MCIN with funding from the European Union NextGenerationEU (PRTR-C17.I1) promoted by the Government of Aragon. J.C. acknowledges grant CEX2023-001286-S funded by MICIU/AEI /10.13039/501100011033 and grant PID2023-148359NB-C21 funded by MICIU/AEI /10.13039/501100011033 and the European Union FEDER.
\end{acknowledgments}

\appendix

\section{Accuracy of the NEP potential}
\label{subsec:NEP-accuracy}

\begin{figure}[htbp]
    \centering
    \includegraphics[width=1\linewidth]{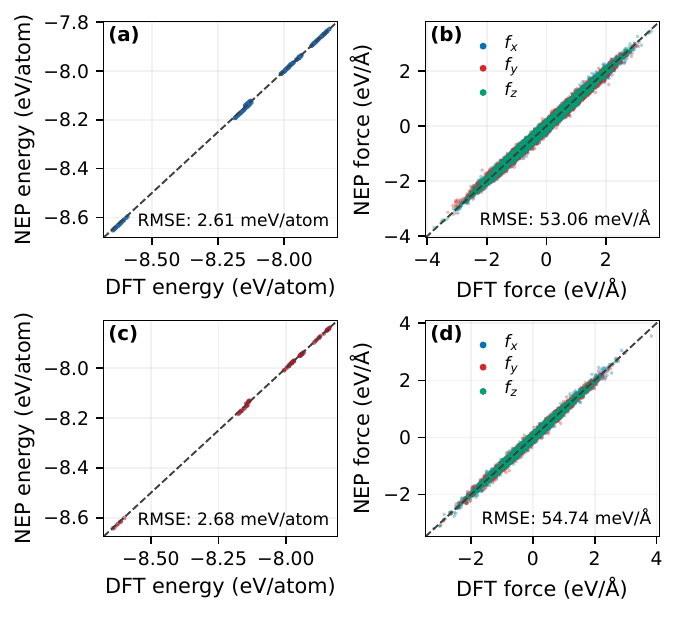}
    \caption{NEP predictions compared with the DFT reference for the training and test sets. Panels (a) and (b) show the training-set energies and forces, respectively; panels (c) and (d) show the corresponding test-set results. Dashed lines indicate perfect agreement.}
    \label{fig:RMSE}
\end{figure}

The production NEP was trained on the $820$-structure training set and validated on the $100$-structure test set described in Sec.~\ref{subsec:GPUMD}. The final potential reaches root-mean-square errors of $2.61$ and \SI{2.68}{\milli\electronvolt\per\atom} in energy and $53.1$ and \SI{54.7}{\milli\electronvolt\per\angstrom} in force on the training and test sets, respectively. The energy and force parity plots in Fig.~\ref{fig:RMSE}, confirming the accuracy of the quantities entering the force and heat-current calculations. This potential is used for all calculations reported below.

\section{Convergence and wave-vector-resolved results}
\label{app:suppfigs}

\begin{figure}[htbp]
    \centering
    \figph{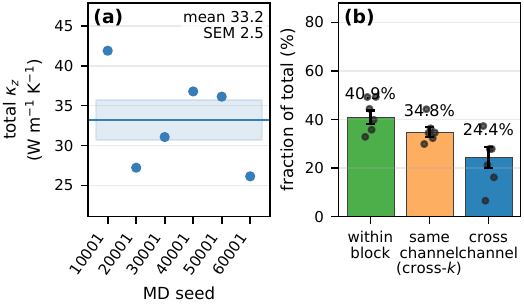}
    \caption{Per-seed stability of the GKMA results at $\SI{300}{K}$ using the $\SI{0.30}{ns}$--$\SI{0.70}{ns}$ correlation-time window. (a) Total axial conductivity for each of the six seeds, with the seed mean ($\SI{33.2}{\watt\per\meter\per\kelvin}$) and standard error ($\SI{2.5}{\watt\per\meter\per\kelvin}$) shown by the line and band. (b) Three-way partition into within-block, same-channel (cross-$k$), and cross-channel contributions. Bars show the ratio-of-sums estimates, error bars are jackknife standard errors, and points are the per-seed fractions.}
    \label{fig:dwnt_gkma_partition}
\end{figure}

\begin{figure}[htbp]
    \centering
    \figph{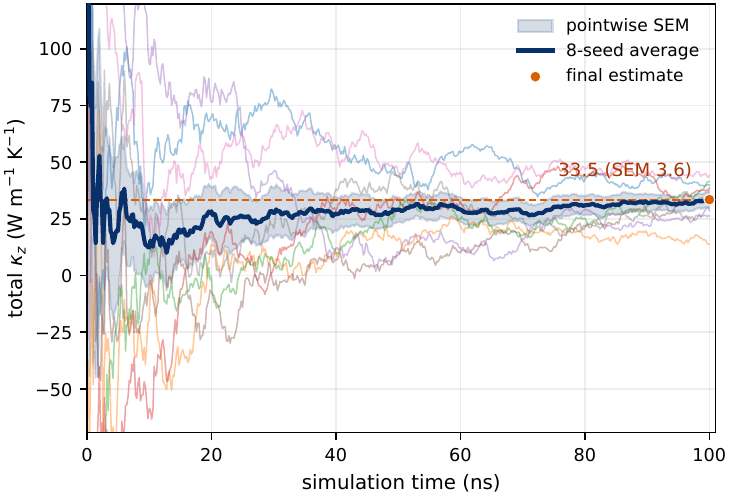}
    \caption{Convergence of the total HNEMA conductivity of the DWNT at $\SI{300}{K}$ and $F_{\mathrm{e}}=\SI{4e-6}{\per\angstrom}$ over the complete $0$--$\SI{100}{ns}$ production trajectory. Thin lines show the cumulative estimator for each of the eight independent seeds; the dark line and shaded band show their mean and pointwise standard error. At $\SI{100}{ns}$, the eight-seed mean is $\SI{33.5}{\watt\per\meter\per\kelvin}$ and the seed standard error is $\SI{3.6}{\watt\per\meter\per\kelvin}$.}
    \label{fig:dwnt_hnema_conv300}
\end{figure}

\begin{figure*}[tbp]
    \centering
    \includegraphics[width=1\textwidth]{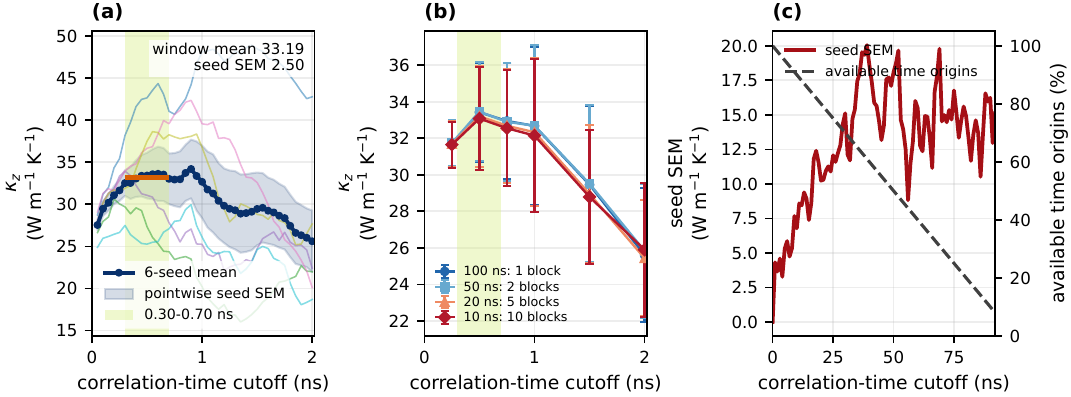}
    \caption{Correlation-time and block-length assessment of the total GKMA axial conductivity of the DWNT at $\SI{300}{K}$. (a) The thin lines show the running conductivity obtained from each complete $\SI{100}{ns}$ trajectory. The blue line and band show the six-seed mean and standard error. The green region marks the first broad plateau of the six-seed mean, from $\SI{0.30}{ns}$ to $\SI{0.70}{ns}$. The window-averaged conductivity is $\SI{33.19\pm2.50}{\watt\per\meter\per\kelvin}$. (b) Block-length test using 100, 50, 20, and $\SI{10}{ns}$ blocks. (c) Growth of the seed standard error at long correlation times together with the fraction of available time origins.}
    \label{fig:dwnt_gkma_conv}
\end{figure*}

\begin{figure}[tbp]
    \centering
    \includegraphics[width=1\linewidth]{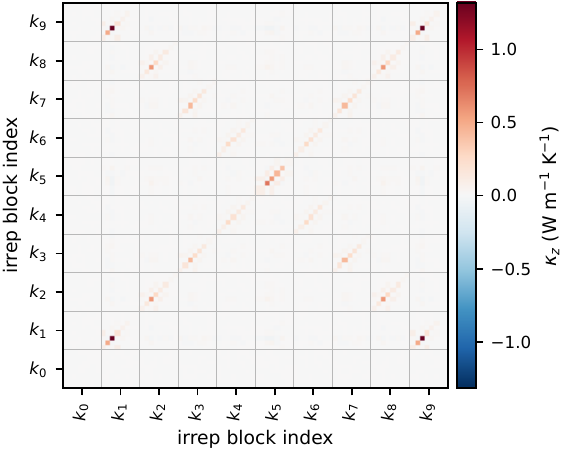}
    \caption{Axial-wave-vector-resolved GKMA irrep-block-pair matrix of the DWNT at $\SI{300}{K}$, indexed by eight $(m,\Pi_{\mathrm{V}})$ symmetry channels at each of ten axial wave vectors $k_0\ldots k_9$. The grey grid separates the $k$ sectors. For each of the six $\SI{100}{ns}$ NVE production trajectories, the matrix is evaluated over the complete trajectory and averaged over correlation-time cutoffs from $\SI{0.30}{ns}$ to $\SI{0.70}{ns}$. The six seed-resolved matrices are then averaged and symmetrized as $\tfrac12(\kappa_{IJ}+\kappa_{JI})$.}
    \label{fig:dwnt_gkma_matrix}
\end{figure}

\begin{figure*}[t]
    \centering
    \includegraphics[width=1\textwidth]{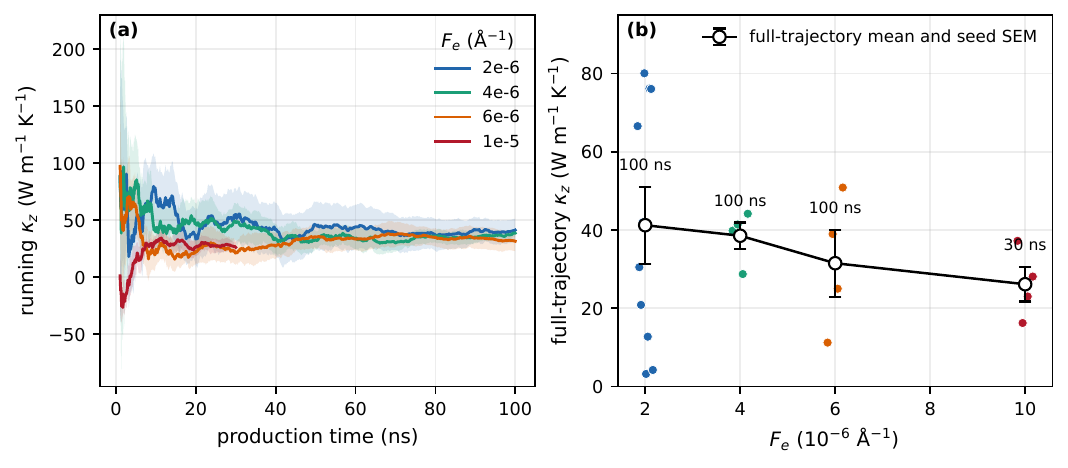}
    \caption{Homogeneous driving-field test for the axial thermal conductivity of the DWNT at $\SI{300}{K}$. (a) Cumulative full-trajectory conductivity for four field magnitudes.  (b) Full-trajectory mean conductivity versus field magnitude. Error bars show their mean and standard error.}
    \label{fig:hnemd_fe_scan}
\end{figure*}

This appendix documents the statistical convergence of the HNEMA and GKMA estimators and retains the full axial-wave-vector resolution underlying the reduced symmetry-channel results. Figure~\ref{fig:dwnt_gkma_partition} shows the per-seed stability of the GKMA total and its three-way correlation partition. Figure~\ref{fig:dwnt_hnema_conv300} shows the seed-resolved HNEMA cumulative estimators and their remaining spread at the end of the trajectories. Figure~\ref{fig:dwnt_gkma_conv} examines the correlation-time and block-length dependence used to select the GKMA averaging window. Figure~\ref{fig:dwnt_gkma_matrix} displays the complete GKMA matrix indexed by the 80 irrep blocks before the axial-wave-vector indices are summed to form the symmetry-channel-pair matrix of Fig.~\ref{fig:dwnt_gkma_channels}. The corresponding single-index HNEMA block contributions are already shown in Fig.~\ref{fig:dwnt_hnema_km}. Finally, Fig.~\ref{fig:hnemd_fe_scan} tests the dependence of the driven conductivity on the homogeneous driving-field magnitude.

\clearpage

\onecolumngrid
\section{Representative symmetry-adapted displacement patterns}
\label{app:irrep-patterns}

\begin{figure}[htbp]
    \centering
    \includegraphics[width=\textwidth]{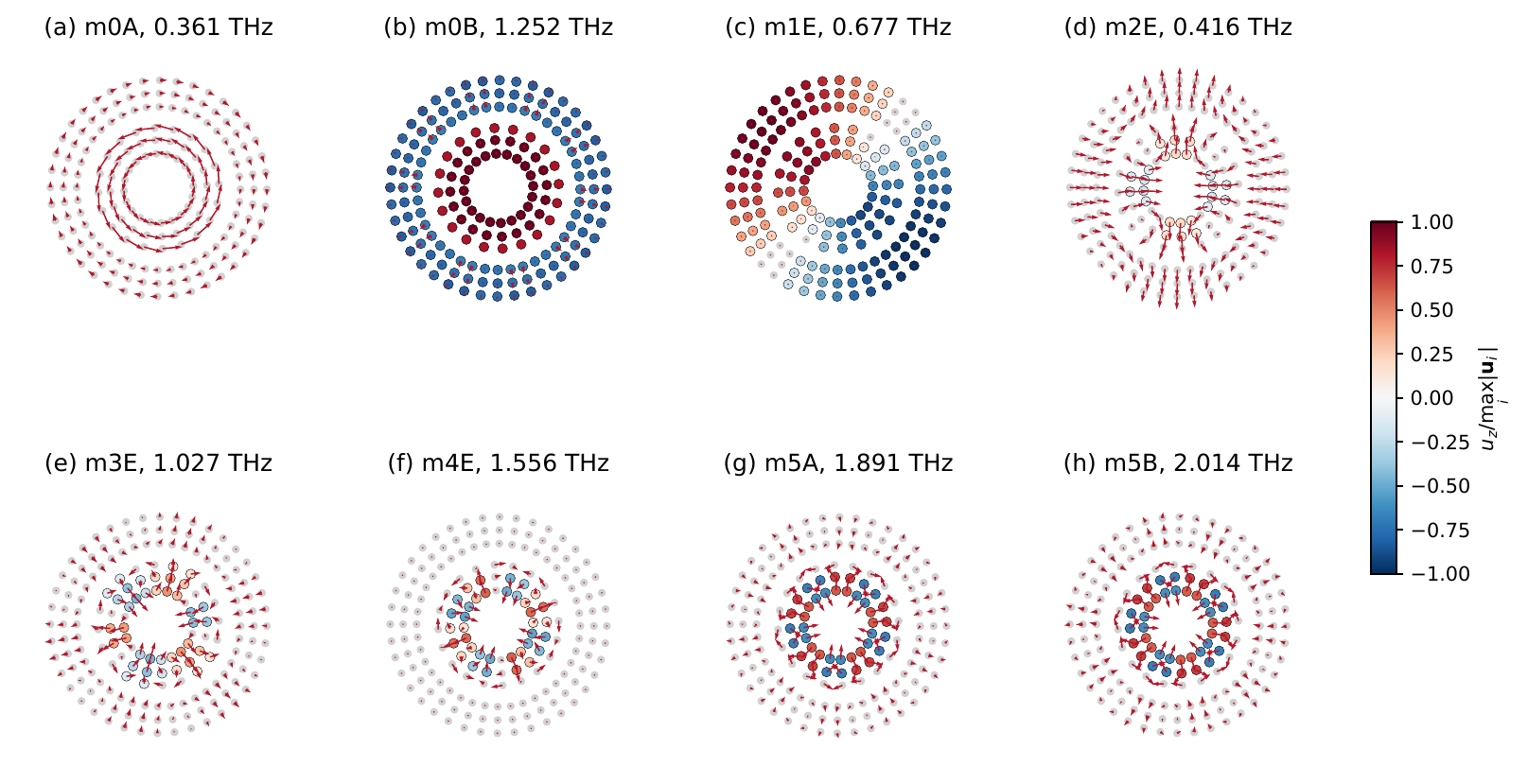}
    \caption{Representative $\Gamma$-point phonon displacement patterns of the DWNT, viewed along the tube axis. Titles give the symmetry channel and frequency. Arrows show transverse displacements; red and blue indicate positive and negative axial displacements. The colorbar gives $u_z/\max_i|\mathbf{u}_i|$, normalized independently for each mode; grey markers denote sites with normalized axial displacement magnitude below 0.12.}
    \label{fig:irrep-patterns}
\end{figure}

Fig.~\ref{fig:irrep-patterns} illustrates representative $\Gamma$-point phonon modes of the DWNT in the SA eigenbasis. For each symmetry channel, we select the lowest-frequency mode above $\SI{0.05}{THz}$, excluding the near-zero-frequency rigid motions. These examples illustrate the symmetry classification rather than the dominant heat-carrying modes.

\clearpage
\twocolumngrid

 \bibliography{apssamp}  

@misc{cen2026pulgontoolstoolkitanalysingharnessing,
      title={Pulgon-tools: A toolkit for analysing and harnessing symmetries in quasi-1D systems}, 
      author={Yu-Jie Cen and Sandro Wieser and Georg K. H. Madsen and Jesús Carrete},
      year={2026},
      eprint={2603.28435},
      archivePrefix={arXiv},
      primaryClass={physics.comp-ph},
      url={https://arxiv.org/abs/2603.28435}, 
}

@Article{Cen2026,
author={Cen, Yu-Jie
and Wieser, Sandro
and Madsen, Georg K. H.
and Carrete, Jes{\'u}s},
title={Ab-initio heat transport in defect-laden quasi-1D systems from a symmetry-adapted perspective},
journal={npj Computational Materials},
year={2026},
month={Jan},
day={09},
volume={12},
number={1},
pages={19},
issn={2057-3960},
doi={10.1038/s41524-025-01866-1},
url={https://doi.org/10.1038/s41524-025-01866-1}
}

@book{damnjanovic2010line,
  title     = {Line Groups in Physics: Theory and Applications to Nanotubes and Polymers},
  author    = {Damnjanovic, M. and Milosevic, I.},
  isbn      = {9783642111716},
  lccn      = {2010921144},
  series    = {Lecture Notes in Physics},
  year      = {2010},
  publisher = {Springer}
}

@article{Lv_2016,
doi = {10.1088/1367-2630/18/1/013028},
url = {https://doi.org/10.1088/1367-2630/18/1/013028},
year = {2016},
month = {jan},
publisher = {IOP Publishing},
volume = {18},
number = {1},
pages = {013028},
author = {Lv, Wei and Henry, Asegun},
title = {Direct calculation of modal contributions to thermal conductivity via Green–Kubo modal analysis},
journal = {New Journal of Physics},
}

@article{10.1063/1.5081722,
    author = {Seyf, Hamid Reza and Gordiz, Kiarash and DeAngelis, Freddy and Henry, Asegun},
    title = {Using Green-Kubo modal analysis (GKMA) and interface conductance modal analysis (ICMA) to study phonon transport with molecular dynamics},
    journal = {Journal of Applied Physics},
    volume = {125},
    number = {8},
    pages = {081101},
    year = {2019},
    month = {02},
    issn = {0021-8979},
    doi = {10.1063/1.5081722},
    url = {https://doi.org/10.1063/1.5081722},
}

@article{EVANS1982457,
title = {Homogeneous NEMD algorithm for thermal conductivity—Application of non-canonical linear response theory},
journal = {Physics Letters A},
volume = {91},
number = {9},
pages = {457-460},
year = {1982},
issn = {0375-9601},
doi = {https://doi.org/10.1016/0375-9601(82)90748-4},
url = {https://www.sciencedirect.com/science/article/pii/0375960182907484},
author = {Denis J. Evans},
}

@article{PhysRevB.103.205421,
  title = {Spectral decomposition of thermal conductivity: Comparing velocity decomposition methods in homogeneous molecular dynamics simulations},
  author = {Gabourie, Alexander J. and Fan, Zheyong and Ala-Nissila, Tapio and Pop, Eric},
  journal = {Phys. Rev. B},
  volume = {103},
  issue = {20},
  pages = {205421},
  numpages = {11},
  year = {2021},
  month = {May},
  publisher = {American Physical Society},
  doi = {10.1103/PhysRevB.103.205421},
  url = {https://link.aps.org/doi/10.1103/PhysRevB.103.205421}
}

@article{10.1063/1.4921108,
    author = {Feng, Tianli and Qiu, Bo and Ruan, Xiulin},
    title = {Anharmonicity and necessity of phonon eigenvectors in the phonon normal mode analysis},
    journal = {Journal of Applied Physics},
    volume = {117},
    number = {19},
    pages = {195102},
    year = {2015},
    month = {05},
    issn = {0021-8979},
    doi = {10.1063/1.4921108},
    url = {https://doi.org/10.1063/1.4921108},
}

@article{https://doi.org/10.1002/mgea.70028,
author = {Xu, Ke and Bu, Hekai and Pan, Shuning and Lindgren, Eric and Wu, Yongchao and Wang, Yong and Liu, Jiahui and Song, Keke and Xu, Bin and Li, Yifan and Hainer, Tobias and Svensson, Lucas and Wiktor, Julia and Zhao, Rui and Huang, Hongfu and Qian, Cheng and Zhang, Shuo and Zeng, Zezhu and Zhang, Bohan and Tang, Benrui and Xiao, Yang and Yan, Zihan and Shi, Jiuyang and Liang, Zhixin and Wang, Junjie and Liang, Ting and Cao, Shuo and Wang, Yanzhou and Ying, Penghua and Xu, Nan and Chen, Chengbing and Zhang, Yuwen and Chen, Zherui and Wu, Xin and Jiang, Wenwu and Berger, Esme and Li, Yanlong and Chen, Shunda and Gabourie, Alexander J. and Dong, Haikuan and Xiong, Shiyun and Wei, Ning and Chen, Yue and Xu, Jianbin and Ding, Feng and Sun, Zhimei and Ala-Nissila, Tapio and Harju, Ari and Zheng, Jincheng and Guan, Pengfei and Erhart, Paul and Sun, Jian and Ouyang, Wengen and Su, Yanjing and Fan, Zheyong},
title = {GPUMD 4.0: A high-performance molecular dynamics package for versatile materials simulations with machine-learned potentials},
journal = {Materials Genome Engineering Advances},
volume = {3},
number = {3},
pages = {e70028},
doi = {https://doi.org/10.1002/mgea.70028},
url = {https://onlinelibrary.wiley.com/doi/abs/10.1002/mgea.70028},
year = {2025}
}

@book{wigner2012group,
  title     = {Group theory: and its application to the quantum mechanics of atomic spectra},
  author    = {Wigner, Eugene},
  volume    = {5},
  year      = {2012},
  publisher = {Elsevier}
}

@article{M_Damnjanovic_1994,
  doi       = {10.1088/0305-4470/27/14/014},
  year      = {1994},
  month     = {jul},
  publisher = {},
  volume    = {27},
  number    = {14},
  pages     = {4859},
  author    = {M Damnjanovic and  I Milosevic},
  title     = {Modified group-projector technique: subgroups and generators},
  journal   = {J. Phys. A: Math. Gen.}
}

@article{DAMNJANOVIC20151,
  title    = {Full symmetry implementation in condensed matter and molecular {physics—Modified} group projector technique},
  journal  = {Phys. Rep.},
  volume   = {581},
  pages    = {1-43},
  year     = {2015},
  issn     = {0370-1573},
  doi      = {10.1016/j.physrep.2015.04.002},
  author   = {Milan Damnjanović and Ivanka Milošević}
}

@article{PhysRevB.50.17953,
  title = {Projector augmented-wave method},
  author = {Bl\"ochl, P. E.},
  journal = {Phys. Rev. B},
  volume = {50},
  issue = {24},
  pages = {17953--17979},
  year = {1994},
  doi = {10.1103/PhysRevB.50.17953}
}

@article{PhysRevB.54.11169,
  title = {Efficient iterative schemes for \textit{ab initio} total-energy calculations using a plane-wave basis set},
  author = {Kresse, G. and Furthm\"uller, J.},
  journal = {Phys. Rev. B},
  volume = {54},
  issue = {16},
  pages = {11169--11186},
  year = {1996},
  doi = {10.1103/PhysRevB.54.11169}
}

@article{KRESSE199615,
  title = {Efficiency of ab-initio total energy calculations for metals and semiconductors using a plane-wave basis set},
  author = {Kresse, G. and Furthm\"uller, J.},
  journal = {Comput. Mater. Sci.},
  volume = {6},
  number = {1},
  pages = {15--50},
  year = {1996},
  doi = {10.1016/0927-0256(96)00008-0}
}

@article{PhysRevB.59.1758,
  title = {From ultrasoft pseudopotentials to the projector augmented-wave method},
  author = {Kresse, G. and Joubert, D.},
  journal = {Phys. Rev. B},
  volume = {59},
  issue = {3},
  pages = {1758--1775},
  year = {1999},
  doi = {10.1103/PhysRevB.59.1758}
}

@article{PhysRevLett.100.136406,
  title = {Restoring the Density-Gradient Expansion for Exchange in Solids and Surfaces},
  author = {Perdew, John P. and Ruzsinszky, Adrienn and Csonka, G\'abor I. and Vydrov, Oleg A. and Scuseria, Gustavo E. and Constantin, Lucian A. and Zhou, Xiaolan and Burke, Kieron},
  journal = {Phys. Rev. Lett.},
  volume = {100},
  issue = {13},
  pages = {136406},
  year = {2008},
  doi = {10.1103/PhysRevLett.100.136406}
}

@article{10.1002/jcc.21759,
  title = {Effect of the damping function in dispersion corrected density functional theory},
  author = {Grimme, Stefan and Ehrlich, Stephan and Goerigk, Lars},
  journal = {J. Comput. Chem.},
  volume = {32},
  number = {7},
  pages = {1456--1465},
  year = {2011},
  doi = {10.1002/jcc.21759}
}

@misc{kaplan2025foundationalpotentialenergysurface,
      title={A Foundational Potential Energy Surface Dataset for Materials}, 
      author={Aaron D. Kaplan and Runze Liu and Ji Qi and Tsz Wai Ko and Bowen Deng and Janosh Riebesell and Gerbrand Ceder and Kristin A. Persson and Shyue Ping Ong},
      year={2025},
      eprint={2503.04070},
      archivePrefix={arXiv},
      primaryClass={cond-mat.mtrl-sci},
      url={https://arxiv.org/abs/2503.04070}, 
}

@article{10.1063/5.0297006,
    author = {Batatia, Ilyes and Benner, Philipp and Chiang, Yuan and Elena, Alin M. and Kovács, Dávid P. and Riebesell, Janosh and Advincula, Xavier R. and Asta, Mark and Avaylon, Matthew and Baldwin, William J. and Berger, Fabian and Bernstein, Noam and Bhowmik, Arghya and Bigi, Filippo and Blau, Samuel M. and Cărare, Vlad and Ceriotti, Michele and Chong, Sanggyu and Darby, James P. and De, Sandip and Della Pia, Flaviano and Deringer, Volker L. and Elijošius, Rokas and El-Machachi, Zakariya and Fako, Edvin and Falcioni, Fabio and Ferrari, Andrea C. and Gardner, John L. A. and Gawkowski, Mikołaj J. and Genreith-Schriever, Annalena and George, Janine and Goodall, Rhys E. A. and Grandel, Jonas and Grey, Clare P. and Grigorev, Petr and Han, Shuang and Handley, Will and Heenen, Hendrik H. and Hermansson, Kersti and Ho, Cheuk Hin and Hofmann, Stephan and Holm, Christian and Jaafar, Jad and Jakob, Konstantin S. and Jung, Hyunwook and Kapil, Venkat and Kaplan, Aaron D. and Karimitari, Nima and Kermode, James R. and Kourtis, Panagiotis and Kroupa, Namu and Kullgren, Jolla and Kuner, Matthew C. and Kuryla, Domantas and Liepuoniute, Guoda and Lin, Chen and Margraf, Johannes T. and Magdău, Ioan-Bogdan and Michaelides, Angelos and Moore, J. Harry and Naik, Aakash A. and Niblett, Samuel P. and Norwood, Sam Walton and O’Neill, Niamh and Ortner, Christoph and Persson, Kristin A. and Reuter, Karsten and Rosen, Andrew S. and Rosset, Louise A. M. and Schaaf, Lars L. and Schran, Christoph and Shi, Benjamin X. and Sivonxay, Eric and Stenczel, Tamás K. and Sutton, Christopher and Svahn, Viktor and Swinburne, Thomas D. and Tilly, Jules and van der Oord, Cas and Vargas, Santiago and Varga-Umbrich, Eszter and Vegge, Tejs and Vondrák, Martin and Wang, Yangshuai and Witt, William C. and Wolf, Thomas and Zills, Fabian and Csányi, Gábor},
    title = {A foundation model for atomistic materials chemistry},
    journal = {The Journal of Chemical Physics},
    volume = {163},
    number = {18},
    pages = {184110},
    year = {2025},
    month = {11},
    issn = {0021-9606},
    doi = {10.1063/5.0297006},
    url = {https://doi.org/10.1063/5.0297006},
}

@book{Dresselhaus2008GroupTheory,
  author    = {Dresselhaus, Mildred S. and Dresselhaus, Gene and Jorio, Ado},
  title     = {Group Theory: Application to the Physics of Condensed Matter},
  publisher = {Springer},
  address   = {Berlin, Heidelberg},
  year      = {2008},
  isbn      = {978-3-540-32897-1},
  doi       = {10.1007/978-3-540-32899-5}
}

@article{PhysRevB.103.184302,
  title = {Crystal symmetry based selection rules for anharmonic phonon-phonon scattering from a group theory formalism},
  author = {Yang, Runqing and Yue, Shengying and Quan, Yujie and Liao, Bolin},
  journal = {Phys. Rev. B},
  volume = {103},
  issue = {18},
  pages = {184302},
  numpages = {11},
  year = {2021},
  month = {May},
  publisher = {American Physical Society},
  doi = {10.1103/PhysRevB.103.184302},
  url = {https://link.aps.org/doi/10.1103/PhysRevB.103.184302}
}

@article{phonopy-phono3py-JPCM,
  author  = {Togo, Atsushi and Chaput, Laurent and Tadano, Terumasa and Tanaka, Isao},
  title   = {Implementation strategies in phonopy and phono3py},
  journal = {J. Phys. Condens. Matter},
  volume  = {35},
  number  = {35},
  pages   = {353001},
  year    = {2023},
  doi     = {10.1088/1361-648X/acd831}
}

@article{phonopy-phono3py-JPSJ,
  author  = {Togo, Atsushi},
  title   = {First-principles Phonon Calculations with Phonopy and Phono3py},
  journal = {J. Phys. Soc. Jpn.},
  volume  = {92},
  number  = {1},
  pages   = {012001},
  year    = {2023},
  doi     = {10.7566/JPSJ.92.012001}
}

@book{BradleyCracknell2009,
author    = {Bradley, C. J. and Cracknell, A. P.},
title     = {The Mathematical Theory of Symmetry in Solids:
            Representation Theory for Point Groups and Space Groups},
publisher = {Oxford University Press},
year      = {2009},
doi       = {10.1093/oso/9780199582587.001.0001},
isbn      = {9780199582587},
     }

@article{Carrete2019,
    author  = {Carrete, Jes{\'u}s and Ngoc Tuoc, Vu and Madsen, Georg K. H.},
    title   = {Using nanotubes to study the phonon spectrum of two-dimensional materials},
    journal = {Phys. Chem. Chem. Phys.},
    year    = {2019},
    volume  = {21},
    pages   = {5215--5223},
    doi     = {10.1039/C9CP00052F},
  }

@article{Lin2022,
author  = {Lin, Changpeng and Ponc{\'e}, Samuel and Marzari, Nicola},
title   = {General invariance and equilibrium conditions for lattice dynamics in {1D}, {2D}, and {3D}
materials},
journal = {npj Comput. Mater.},
year    = {2022},
volume  = {8},
pages   = {236},
doi     = {10.1038/s41524-022-00920-6},
}

@article{Lindsay2016,
    author  = {Lindsay, Lucas},
    title   = {First Principles {Peierls--Boltzmann} Phonon Thermal Transport: A Topical Review},
    journal = {Nanoscale Microscale Thermophys. Eng.},
    year    = {2016},
    volume  = {20},
    number  = {2},
    pages   = {67--84},
    doi     = {10.1080/15567265.2016.1218576},
  }

@article{OngZhang2015,
    author  = {Ong, Zhun-Yong and Zhang, Gang},
    title   = {Efficient approach for modeling phonon transmission probability in nanoscale interfacial
    thermal transport},
    journal = {Phys. Rev. B},
    year    = {2015},
    volume  = {91},
    pages   = {174302},
    doi     = {10.1103/PhysRevB.91.174302},
  }

@article{Ong2018,
    author  = {Ong, Zhun-Yong},
    title   = {Tutorial: Concepts and numerical techniques for modeling individual phonon transmission at
    interfaces},
    journal = {J. Appl. Phys.},
    year    = {2018},
    volume  = {124},
    pages   = {151101},
    doi     = {10.1063/1.5048234},
  }

@article{Lindsay2009,
    author  = {Lindsay, L. and Broido, D. A. and Mingo, Natalio},
    title   = {Lattice thermal conductivity of single-walled carbon nanotubes: Beyond the relaxation time
    approximation and phonon--phonon scattering selection rules},
    journal = {Phys. Rev. B},
    year    = {2009},
    volume  = {80},
    pages   = {125407},
    doi     = {10.1103/PhysRevB.80.125407},
  }

@article{McGaughey2019,
author  = {McGaughey, Alan J. H. and Jain, Ankit and
           Kim, Hyun-Young and Fu, Bo},
title   = {Phonon properties and thermal conductivity from first
           principles, lattice dynamics, and the {Boltzmann}
           transport equation},
journal = {Journal of Applied Physics},
volume  = {125},
number  = {1},
pages   = {011101},
year    = {2019},
doi     = {10.1063/1.5064602}
}

@Article{Ercole2016,
author={Ercole, Loris
and Marcolongo, Aris
and Umari, Paolo
and Baroni, Stefano},
title={Gauge Invariance of Thermal Transport Coefficients},
journal={Journal of Low Temperature Physics},
year={2016},
month={Oct},
day={01},
volume={185},
number={1},
pages={79-86},
issn={1573-7357},
doi={10.1007/s10909-016-1617-6},
url={https://doi.org/10.1007/s10909-016-1617-6}
}

\end{document}